\documentclass{emulateapj}

\slugcomment{To appear in The Astrophysical Journal}

\shortauthors{HALPERN ET AL.}
\shorttitle{OUTBURST OF AXP 1E~1547.0$-$5408}

\begin{document}


\def\mag{1E~1547.0$-$5408}
\def\psr{PSR~J1550$-$5418}
\def\snr{G327.24$-$0.13}
\def\xte{XTE~J1810$-$197}
\def\asca{{\em ASCA\/}}
\def\chandra{{\em Chandra\/}}
\def\einstein{{\em Einstein\/}}
\def\swift{{\em Swift\/}}
\def\xmm{{\em XMM-Newton\/}}

\title{Outburst of the 2 s Anomalous X-ray Pulsar 1E~1547.0--5408}

\author{J.~P. Halpern,\altaffilmark{1}
  E.~V. Gotthelf,\altaffilmark{1}
  J.~Reynolds,\altaffilmark{2}
  S.~M.~Ransom,\altaffilmark{3}
  F.~Camilo\altaffilmark{1}
}

\altaffiltext{1}{Columbia Astrophysics Laboratory, Columbia University,
  New York, NY 10027.}
\altaffiltext{2}{Australia Telescope National Facility, CSIRO, Parkes
  Observatory, Parkes, NSW 2870, Australia.}
\altaffiltext{3}{National Radio Astronomy Observatory, Charlottesville, 
  VA 22903.}

\begin{abstract}
Following our discovery of radio pulsations from the newly
recognized Anomalous X-ray Pulsar (AXP) \mag, we initiated
X-ray monitoring with the \swift\ X-ray Telescope,
and obtained a single target-of-opportunity
observation with the {\em Newton X-ray Multi-Mirror Mission}
(\xmm).  In comparison with its historic minimum flux of
$3 \times 10^{-13}$ ergs cm$^{-2}$ s$^{-1}$,
the source was found to be in a record high state,
$f_X(1$--$8\,\mbox{keV}) = 5 \times 10^{-12}$ ergs cm$^{-2}$ s$^{-1}$,
or $L_X = 1.7 \times 10^{35}(d/9\ {\rm kpc})^2$ ergs~s$^{-1}$,
and declining by 25\% in 1 month.
Extrapolating the decay, we bound the total energy in this outburst
to $10^{42} < E < 10^{43}$ ergs.
The spectra (fitted with a Comptonized blackbody)
show that an increase in the temperature and area
of a hot region, to 0.5~keV and $\sim 16\%$
of the surface area of the neutron star, respectively,
are primarily
responsible for its increase in luminosity.
The energy, spectrum, and timescale of decay are consistent with
a deep crustal heating event, similar to an interpretation of the
X-ray turn-on of the transient AXP \xte.
Simultaneous with the 4.6 hour \xmm\ observation,
we observed at 6.4~GHz with the Parkes telescope,
measuring the phase relationship of the radio and X-ray pulse.
The X-ray pulsed fraction of \mag\ is only $\sim 7\%$,
while its radio
pulse is relatively broad for such a slow pulsar,
which may indicate a nearly aligned rotator.
As also inferred from the transient behavior of \xte, the only other AXP
known to emit in the radio, the magnetic field rearrangement responsible
for this X-ray outburst of \mag\ is probably the cause of its radio
turn-on.

\end{abstract}

\keywords{ISM: individual (G327.24--0.13) --- pulsars: individual
(1E~1547.0--5408, PSR~J1550--5418, XTE~J1810--197) --- stars: neutron}

\section{Introduction}\label{sec:intro} 

Anomalous X-ray pulsars (AXPs) and soft gamma-ray repeaters (SGRs)
are young neutron stars (NSs) with rotation period of 2--12\,s and inferred
surface magnetic field strength $B \approx 10^{14-15}$\,G.  See
\citet{wt06} and \citet{kas07} for recent reviews.  In the magnetar model
\citep[][1996]{dt92a,td95}\nocite{td96a}, the rearrangement and decay
of their extreme fields is responsible for their large and variable X-ray
luminosity, which exceeds that available from rotational braking.
Thirteen magnetars are confirmed\footnote{Nine
AXPs and four SGRs; there are two more candidates.  See catalog at
http://www.physics.mcgill.ca/$\sim$pulsar/magnetar/main.html.}, of which
the AXP \mag\ is the most recent.
Discovered with the \einstein\ X-ray satellite in 1980 \citep{lam81},
\mag\ was
identified as a magnetar candidate in the center of the small candidate
SNR \snr\ by \citet{gg07}.  It was subsequently observed to be emitting radio
pulsations at a period of $P=2$~s \citep{cam07b}, with spin-down properties
of a magnetar ($\dot P = 2.3 \times 10^{-11}, B = 2.2 \times 10^{14}$~G),
and a distance estimate of 9~kpc from its dispersion measure (DM).
\mag\ (\psr) and the 5.5\,s AXP \xte\ are the only magnetars known to emit in
the radio \citep{crh+06}.  Both are demonstrably transient radio sources,
having not been detected in previous surveys of adequate sensitivity.

Here, we report on new X-ray observations of \mag\ made shortly after
the radio discovery of its pulsations,
and compare them with archival X-ray data
in order to understand the nature of its apparent X-ray outburst
from its spectral and pulse properties.
Radio pulsar coverage was also obtained at the same time as one of the
X-ray observations, and the phase relationship of the radio and X-ray
pulse was determined.

\section{X-ray Observations}\label{sec:obs}

\subsection{Swift}

\begin{figure}[t]
\begin{center}
\includegraphics[angle=0,scale=0.45]{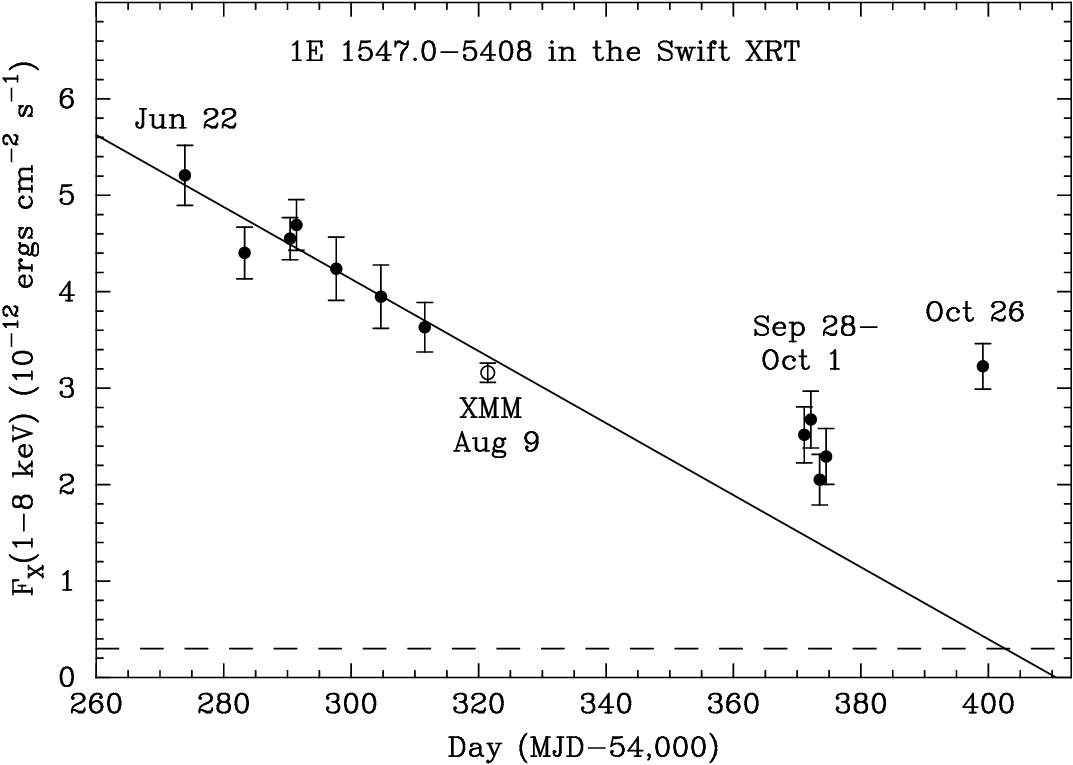}
\caption{\label{fig:decay}
Absorbed, $1-8$~keV
X-ray flux of \mag\ beginning on 2007 June 22 from the \swift\ XRT
({\it filled circles}) and including \xmm\ on 2007 August 9
({\it open circle}).  The solid line is fitted only to the early
\swift\ points.  The dashed line indicates the minimum historical flux
level, which was seen by \chandra\ and \xmm\ in 2006 July and August
\citep[][and Table~\ref{tab:xrayspec}]{gg07}.
}
\end{center}
\end{figure}

Following the detection of radio pulsations from \mag\ on 2007 June 8,
seven observations were made with the \swift\
X-ray telescope \citep[XRT;][]{geh04,bur05} between 2007 June 22 and July 30.
The data were taken in photon counting mode, which provides imaging with
$18''$ half-power diameter resolution and 2.5\,s time sampling.
Neither the counts nor the time resolution
were sufficient to detect pulsations.
We used the archived event files
with standard screening criteria applied.
A total of 22.4~ks of exposure was obtained, although
2.6~ks of this had incorrect attitude solution; the latter was not
used for spectral analysis.
The source was detected in the range 1--8\,keV with an initial count
rate of 0.095\ s$^{-1}$, and a systematic decrease of about 30\%
over the 40 day interval (see Fig.~\ref{fig:decay}).
The seven \swift\ observations were summed in order to accumulate
sufficient statistics for a useful spectral fit.
The spectral modeling is discussed in \S\ref{sec:spectra}.

The maximum observed flux,
$\approx 5.2 \times 10^{-12}$\,ergs\,cm$^{-2}$\,s$^{-1}$ in the
1--8~keV band on 2007 June 22, is 16 times greater
than the minimum seen in 2006 July--August, and 2.6 times greater than
the previous highest flux \citep{gg07}. 
The small dynamic range of the data in Figure~\ref{fig:decay}
prevents a characterization of the decay curve, whether
linear, exponential, or power law.
Additional observations were made by \swift\
on 2007 September 28 -- October 1, after a 2 month gap caused by
a gyroscope anomaly, and finally on October 26.
These show the flux leveling off and
possibly rising again.  The proximity
of the Sun prevents further observations until early 2008.

\subsection{XMM-Newton}

The results of all archival X-ray observations of \mag\
were published by \citet{gg07}.  Here, we reanalyze their
long (46~ks) \xmm\ observation of 2006 August 21, which found
the source in its lowest observed state,
in order to compare the components of its
X-ray spectrum with the high state and to make a
more sensitive search for pulsations at the now known period.
We reprocessed the observation data files using the {\tt emchain}
and {\tt epchain} scripts under Science Analysis System (SAS) version
xmmsas\_20060628\_1801-7.0.0. The 73.4~ms sampling of the 
European Photon Imaging Camera pn CCD \citep[EPIC pn;][]{tur03}
in ``full frame'' mode was adequate to search for pulsations from \mag.
The 0.9~s sampling of the two EPIC MOS cameras,
operated in ``large window'' mode, is close to the
Nyquist frequency for the 2~s pulsar.

On 2007 August 9 we obtained a new \xmm\ observation.
The exposure time was 4.2\,hr in the EPIC pn and 4.6\,hr
in the EPIC MOS. The pn CCD was
operated in ``large window'' mode, with 48~ms sampling.
The two MOS CCDs were operated in ``small window''
mode, with time sampling of 0.3~s, in order to be used to
study the pulsed light curve.  We processed the data using
the SAS version mentioned above.  This observation came 10 days
after a string of \swift\ observations, and its flux falls
on the linear decay fitted to the prior \swift\ points
(see Fig.~\ref{fig:decay}).  Spectral fitting
of the \xmm\ observations is discussed in \S\ref{sec:spectra}
and timing analysis for pulsations in \S\ref{sec:pulse}.

\section{X-ray Spectral Fitting}\label{sec:spectra}

We first tried fitting a single blackbody to the spectra
from \xmm\ and \swift.  A single blackbody is an inadequate fit
in either high or low states.  It leaves a deficit
of flux at 1 keV, and an excess at the highest energies, $>5$~keV.
The low-energy end of the spectrum may be affected by an unmodeled
proton cyclotron absorption line, of resonance energy
$E_c = 0.63(1+z)^{-1}(B/10^{14}\,{\rm G})$~keV, where $z$ is the
gravitational redshift and $(1+z) = (1-2GM/c^2R)^{-1/2} \approx 1.3$.
For $B = 2.2 \times 10^{14}$~G, the line would appear at
$\approx 1$~keV.  However, there is not enough coverage of the
continuum at low energies to model it.

Similarly, the high-energy end of the spectrum is poorly characterized.
However, we can fit the high-energy excess with a
simple Comptonization model that introduces only
one additional parameter.  In this model, surface thermal photons
are multiply scattered by relativistic electrons of mean energy
$\gamma mc^2$ and optical depth $\tau_{\rm es}$ \citep{poz76,ryb79}.
The mean energy boost factor per scattering is
$A=4\langle\gamma^2\rangle/3$.
In the limit $\tau_{\rm es}< 1$, an analytic approximation for the
spectrum can be derived that is a function only of the surface blackbody
temperature $T_{\rm BB}$ and the combined scattering parameter
$\alpha = -{\rm ln}\tau_{\rm es}/{\rm ln}A$.
The scattered spectrum resembles the original blackbody, plus a tail that
at high energies approaches a power law of photon index
$\Gamma = 2 + \alpha$.
This model was also used by \cite{tie05} to fit
spectra of another AXP, although they
derived (in our notation) $\Gamma = 1 + \alpha$.

\begin{figure}[t]
\begin{center}
\includegraphics[angle=270,scale=0.35]{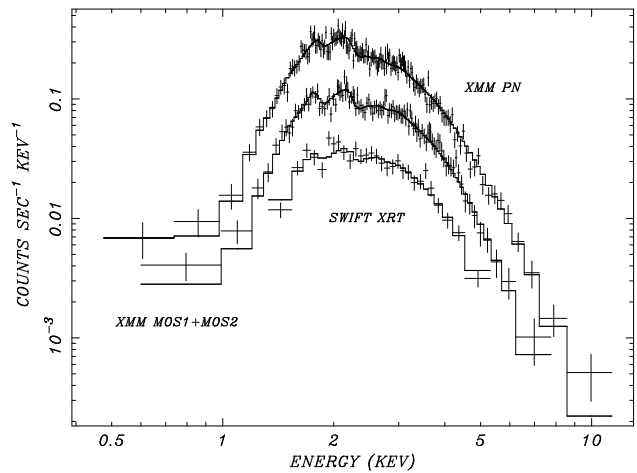}
\caption{\label{fig:xrayspec}
Spectra of \mag\ from the \xmm\ observation of 2007 August 9,
showing EPIC pn and MOS separately,
and the summed \swift\ XRT observations from 2007 June 22 -- July 30.
The fit is to the Comptonized blackbody model described in
the text, with parameters listed in Table~\ref{tab:xrayspec}.
}
\end{center}
\end{figure}

\begin{deluxetable*}{lccc}
\tablecolumns{4}
\tablewidth{320pt}
\tablecaption{Comptonized Blackbody Spectral Fits}
\tablehead{
\colhead{} & \colhead{2006 Aug 21} & \colhead{2007 Jun 22--Jul 30} & \colhead{2007 Aug 9} \\
\colhead{Parameter} & \colhead{{\it XMM-Newton}} & \colhead{{\it Swift} average} &  \colhead{{\it XMM-Newton}}
}
\startdata
$N_{\rm H}$ ($10^{22}$ cm$^{-2}$)\dotfill  & $2.84 \pm 0.60$ & 3.12 (fixed)    & $3.12 \pm 0.23$ \\
$kT_{\rm BB}$ (keV)\dotfill       & $0.40 \pm 0.06$ & $0.50 \pm 0.03$ & $0.52 \pm 0.03$ \\
$\alpha$\dotfill  & $2.10^{+0.76}_{-0.38}$ &  2.93 (fixed)   & $2.93^{+0.97}_{-0.48}$ \\
$A_{\rm BB}$ (cm$^2$)\dotfill & $3.6 \times 10^{11}$ & $17.6 \times 10^{11}$ & $9.6 \times 10^{11}$ \\
$F$ (ergs cm$^{-2}$ s$^{-1}$)\tablenotemark{a}\dotfill & $3.3 \times 10^{-13}$ & $4.6 \times 10^{-12}$ & $3.2 \times 10^{-12}$ \\
$L_{\rm BB}$ ($10^{34}$ ergs s$^{-1}$)\tablenotemark{b}\dotfill & 0.95 & 11.3 & 7.2 \\
$L_{\rm tot}$ ($10^{34}$ ergs s$^{-1}$)\tablenotemark{c}\dotfill & 1.39 & 15.2 & 9.7 \\
$\chi^2_{\nu}({\nu})$\dotfill   & 0.99(117)   & 1.1(27)   & 0.77(252) \\
\enddata
\label{tab:xrayspec}
\tablecomments{\footnotesize Uncertainties are 90\% confidence intervals for three interesting parameters.
Luminosities and areas are computed for $d=9$~kpc, assuming isotropic flux.}
\tablenotetext{a}{\footnotesize Absorbed flux in the $1-10$ keV band.}
\tablenotetext{b}{\footnotesize Unabsorbed, bolometric blackbody luminosity.}
\tablenotetext{c}{\footnotesize Unabsorbed total luminosity including Comptonized component.}
\end{deluxetable*}

While this approach neglects effects of the magnetic field,
which are treated by the more realistic
resonant cyclotron scattering model \citep{lyu06,fer07,rea07a,guv07},
the effects on the spectrum are similar, and our Comptonized
blackbody model is
in fact equivalent to equation (58) of \citet{fer07},
which they used to fit their own Monte Carlo spectral results.
Either scattering model is more
meaningful than the ``traditional'' pure power-law plus blackbody fit,
because there is no physical mechanism that extends such a power law
to photon energies less than the seed blackbody photons.

The results of spectral fitting are summarized in
Table~\ref{tab:xrayspec} and Figure~\ref{fig:xrayspec}.
Acceptable values of $\chi^2$ and reasonably small errors on the
three fitting parameters $N_{\rm H}, kT_{\rm BB}$, and $\alpha$ were
obtained from both \xmm\ observations.  In the case of \swift,
where fewer photons were collected, we could obtain a small
error on $kT_{\rm BB}$, but $\alpha$ was not constrained.
We therefore fixed $N_{\rm H}$ and $\alpha$ in the fitting
of the \swift\
spectrum to their nearly contemporaneous values measured by \xmm\
in 2007.

From the average \swift\ spectrum of 2007 June--July, most of the
flux is fitted by a blackbody whose bolometric luminosity is
$L_{\rm BB} \approx 1.1 \times 10^{35}(d/9\ {\rm kpc})^2$ ergs~s$^{-1}$,
and the corresponding area is
$A_{\rm BB}\approx 1.8 \times 10^{12}(d/9\ {\rm kpc})^2$\,cm$^2$,
or $\sim 14\%$ of the NS surface.   In addition, there is a Compton
scattered component of
$L_s \approx 3.9 \times 10^{34}(d/9\ {\rm kpc})^2$ ergs~s$^{-1}$.
Together, these components account for the total luminosity
$L_{\rm tot} \approx 1.5 \times 10^{35}(d/9\ {\rm kpc})^2$ ergs~s$^{-1}$
in Table~\ref{tab:xrayspec}.  Relative to these averages of
2007 June--July, the maximum luminosity and area recorded by
the first \swift\ observation of 2007 June 22 are $\approx 13\%$ higher, i.e.,
$L_{\rm BB, max} \approx 1.3 \times 10^{35}(d/9\ {\rm kpc})^2$ ergs~s$^{-1}$,
$A_{\rm BB, max}/A_{\rm NS} \approx 0.16$, and
$L_{s,{\rm max}} \approx 4.4 \times 10^{34}(d/9\ {\rm kpc})^2$ ergs~s$^{-1}$.
Smaller values are then obtained from the subsequent \xmm\ observation
on 2007 August 9, which, within the errors, are consistent with a
decrease of the blackbody area at a constant temperature.
We fitted individual \swift\ observations to look
for more subtle changes in spectral shape.
Within the larger uncertainties of the individual
spectra, there is no evidence for spectral changes
from 2007 June 22 through October 26.

The \xmm\ and \swift\ spectra from 2006 August and 2007 June,
respectively, show that a combination
of the increase in the blackbody temperature from
0.4~keV to 0.5~keV and an increase of the area of that
hot region by a factor of 5 is primarily
responsible for the dramatic increase in X-ray flux.
There is no evidence for a
major change in the efficiency of magnetospheric scattering
between the two states, as the scattering parameter $\alpha$ 
remained the same to within errors.  Only $26-32\%$ of the
X-ray luminosity is in the Compton scattered component in
any state, as
can be seen by comparing $L_{\rm tot}$ and $L_{\rm BB}$
in Table~\ref{tab:xrayspec}.

\section{X-ray Pulsations}\label{sec:pulse}

\begin{figure}[t]
\begin{center}
\includegraphics[angle=270,scale=0.35]{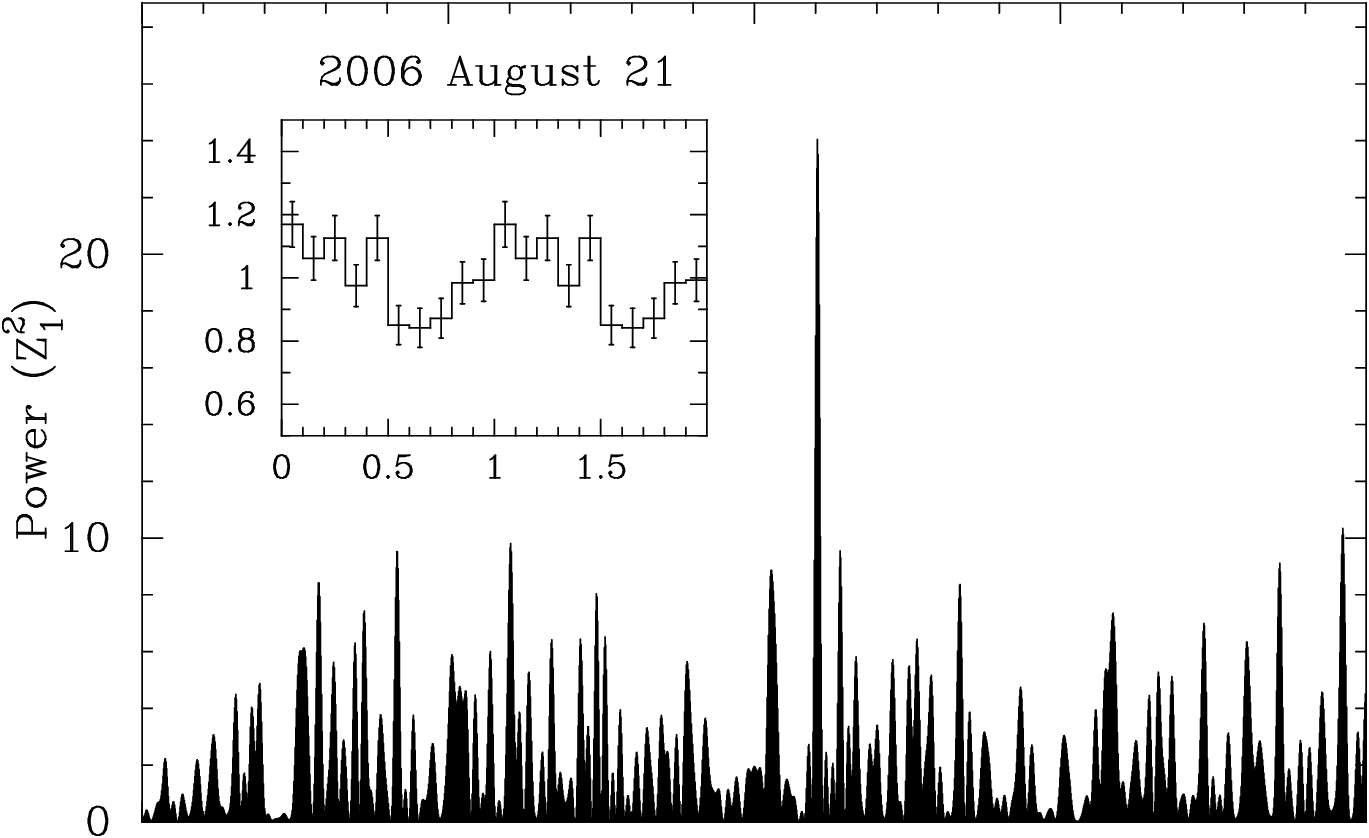}
\includegraphics[angle=270,scale=0.35]{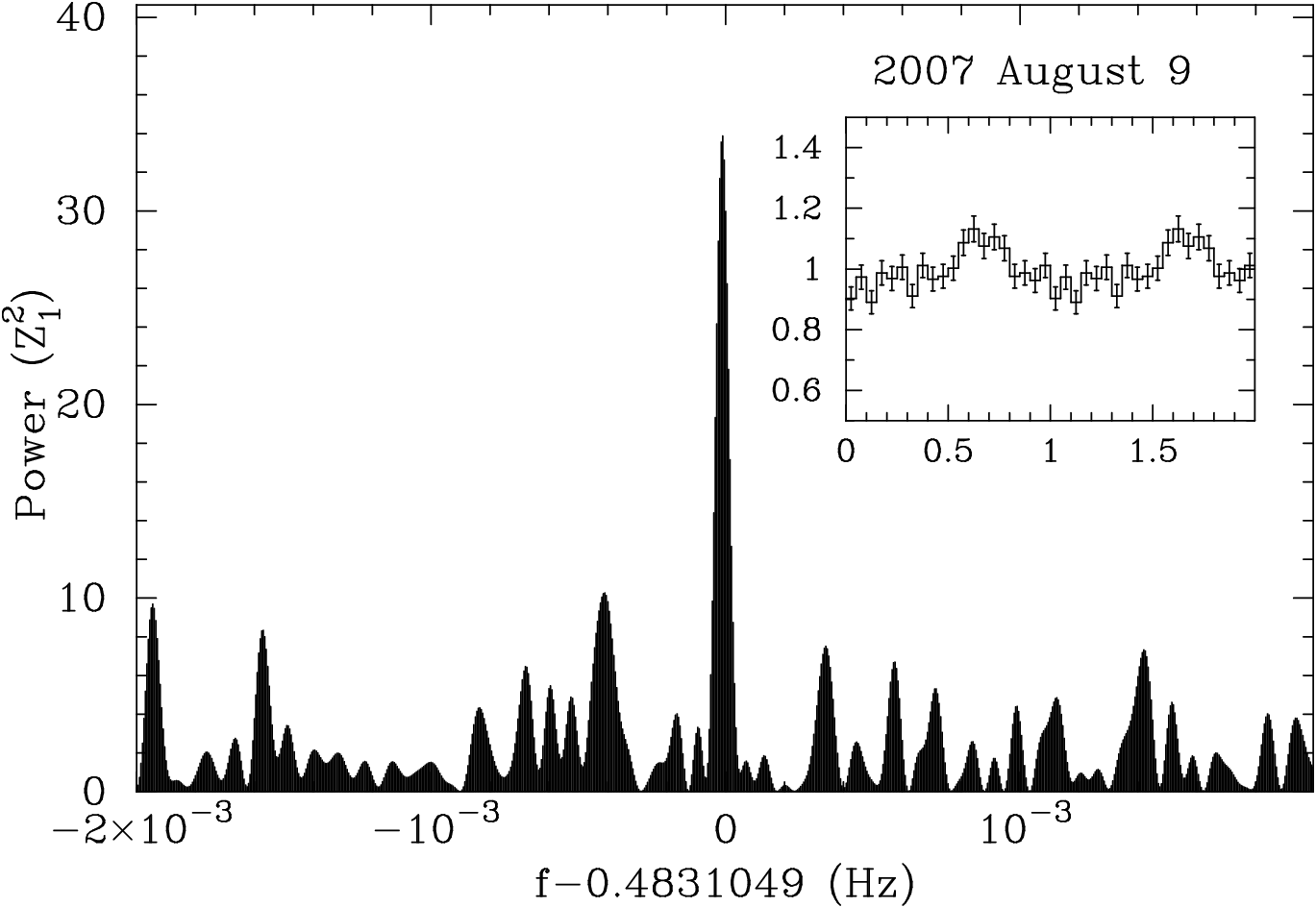}
\caption{\label{fig:power}
$Z_1^2$ power spectra of the combined \xmm\
EPIC pn and MOS observations of \mag.
The reference frequency, 0.4831049 Hz, is the radio measured value on
2007 August 9.  The range of frequencies searched corresponds to
$\dot P \leq 2.8 \times 10^{-10}$ between the two epochs,
or $\le 12$ times the discovery value.
The highest peak on 2006 August 21, if real,
requires an average
$\dot P = 2.8 \times 10^{-11}$ between the two epochs.
{\it Insets\/}: X-ray pulse profiles in the
optimal $1-6$~keV band, normalized to a mean
count rate of 1.  Background has been
subtracted from nearby regions of the CCD.
The 2006 August 21 profile is from the EPIC pn only,
while the 2007 August 9 profile is from the EPIC pn and MOS.
Phase zero of the 2006 August 21 profile is arbitrary.
The 2007 August 9 profile is folded on the
contemporaneous radio ephemeris (see Fig.~\ref{fig:radioxraypulse}).
}
\end{center}
\end{figure}

Only the \xmm\ observations of \mag\ have sufficient photon statistics
and time resolution to search for pulsations.
The event times were converted to Barycentric
Dynamical Time (TDB) using the precise position of the pulsar,
(J2000) R.A. = $15^{\rm h}50^{\rm m}54.11^{\rm s}\pm0.01^{\rm s}$,
Decl. $-54\arcdeg18'23.7''\pm0.1''$
\citep{cam07b}.  Events were extracted from the 2007 August
\xmm\ observation in a $30^{\prime\prime}$ radius around the
source in the EPIC pn CCD, and $25^{\prime\prime}$ in the MOS CCDs.
Photons in the range $0.5-8$~keV from the three detectors were
combined, and searched
for pulsations using the $Z_n^2$ test. A signal was found
consistent with the contemporaneous radio period
with $Z_1^2 = 33.9$, $Z_2^2 = 40.7$, and $Z_3^2 = 43.4$.  
The  $Z_1^2$ power spectrum is shown in Figure~\ref{fig:power}.
These values correspond to a chance probability of $\sim 10^{-7}$.
The power contributed by the harmonics is indicative of a pulse
that is narrower than a sinusoid. 
From the $Z_3^2$ search, we derive $P = 2.069961(17)$~s ($1 \sigma$),
compared with $P = 2.0699348(5)$~s in the radio
(see \S \ref{sec:parkes}).

Figure~\ref{fig:power} also
shows the $1-6$~keV X-ray pulse folded at the exact radio period
(determined in \S \ref{sec:parkes}).
We note that, while the 48~ms
time sampling of the pn is more than sufficient to resolve
the pulse, the 0.3~s sampling of the MOS also adequately
resolves the observed structure.
The X-rays are weakly modulated with pulsed amplitude
$\sim 7\%$, defined as the fraction of total counts above the
minimum of the light curve.
This is small compared to most AXPs, and comparable
only to the small amplitude of 4U~0142+614 \citep{rea07b,gon07}.
The latter has a complex pulse shape, which is energy dependent
as well as varying with luminosity.
Within the limited statistics available for \mag,
we do not have evidence for energy dependence of its pulse shape.

A search of the 2006 August 21 \xmm\ observation of \mag\ at the extrapolated
radio period was less sensitive because (1) the source was in a low
flux state at the time, (2) the 0.9~s sampling of the MOS detectors
in large window mode is close to the Nyquist frequency, and (3) the lack of
a contemporaneous radio ephemeris required a search over several
dozen independent trial periods.  The latter is especially problematic,
as the period derivative of a magnetar can vary by
a factor of several on a timescale of 1 year \citep[e.g.,][]{gav04}.  
Allowing for $0 < \dot P < 2.8 \times 10^{-10}$,
up to 12 times its discovery value of $2.3 \times 10^{-11}$
\citep{cam07b},
requires a search of $\approx 160$ independent periods in
the 2006 August observation of \mag. 

We extracted events from an optimized radius of
$20^{\prime\prime}$ around the source in the 2006 August EPIC pn
image, and applied the $Z_1^2$ test.  The $1-6$~keV
band was chosen to optimize the signal-to-noise.
The strongest signal
in the $\dot P$ search range described above yielded
$Z_1^2=18.1$ and corresponds to $\dot P = 2.8 \times 10^{-11}$,
which is close to its 2007 June value of $\dot P = 2.3 \times 10^{-11}$,
and even closer to the latest value from
continued radio timing, $\dot P = 2.9 \times 10^{-11}$
(F. Camilo et al., in preparation).
The light curve folded at the corresponding period of
2.069067(12)~s is quasi-sinusoidal with a pulsed fraction of
$\approx 15\%$ (see Fig.~\ref{fig:power}).
The overall statistical significance of this
signal is difficult to estimate because the range of
$\dot P$ considered is somewhat arbitrary, but it
is probably 98--99\%.  When we add the source photons 
from the MOS detectors to the search, $Z_1^2$ increases
to 24.0 (see Fig.~\ref{fig:power}),
which is encouraging of the reality of the signal,
as its significance rises to $\approx 99.9\%$ while the period
remains consistent, $P=2.069059(10)$~s.
Therefore, we consider this a tentative detection of pulsations
on 2006 August 21.

\begin{figure}[t]
\begin{center}
\includegraphics[angle=0,scale=0.43]{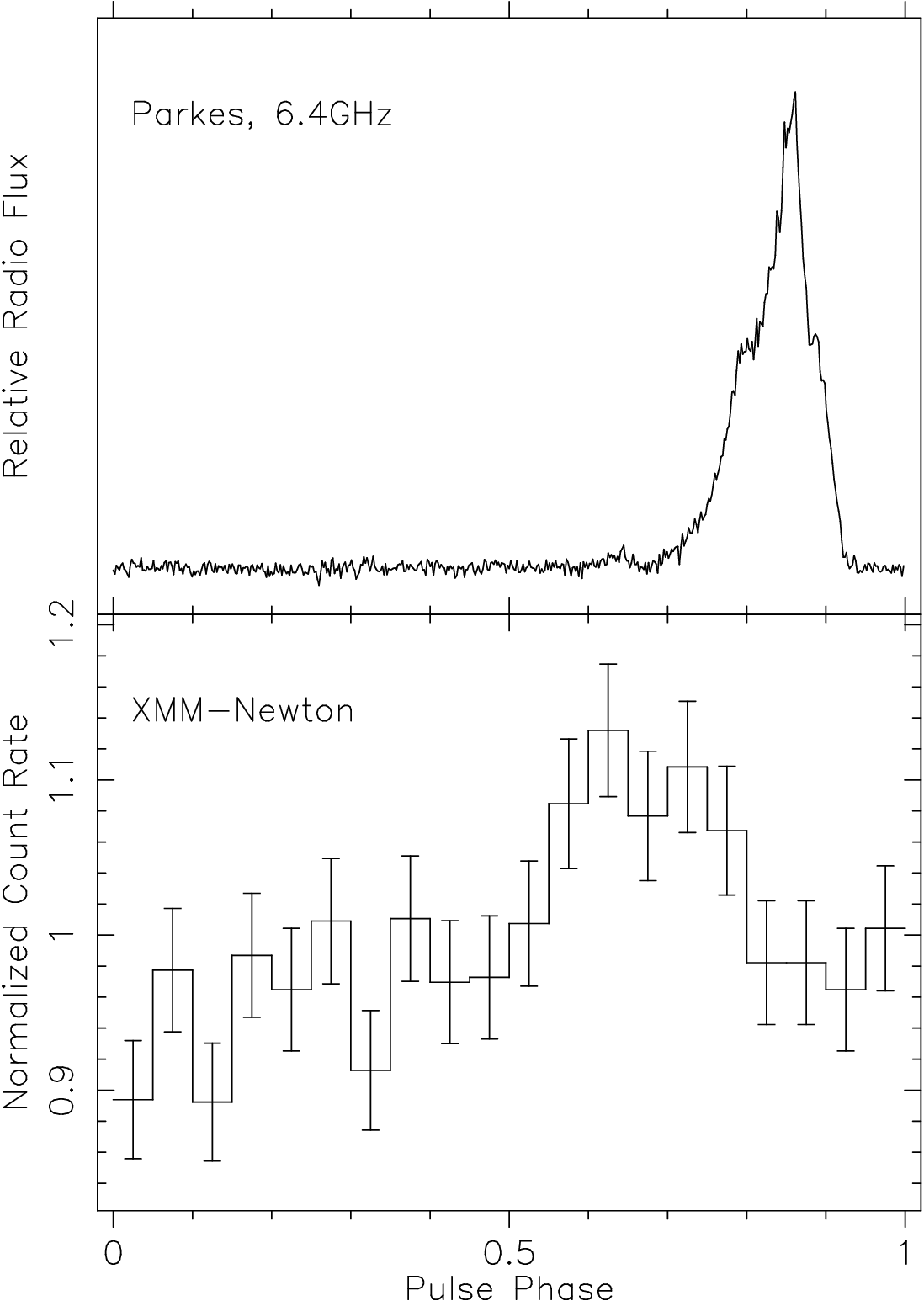}
\caption{\label{fig:radioxraypulse}
Alignment of the simultaneously observed radio
and X-ray pulses on 2007 August 9, folded at
0.4831049 Hz.  Phase zero corresponds to MJD 54321.44 TDB.
The radio (DFB) total intensity profile is displayed in 512 bins, and
the small feature at phase $\approx 0.63$ is real, also being present
in analog filterbank data.
}
\end{center}
\end{figure}
\section{Contemporaneous Radio Pulsar Observation} \label{sec:parkes}

We observed \mag\ with the Parkes telescope in Australia on 2007 August~9,
overlapping with the \xmm\ observation.  We collected data for 4.3\,hr
beginning at 09:57 UT, 16\,min before the start of the X-ray observation.
We used the central beam of the methanol multibeam receiver and recorded
data using two different spectrometers.  With an analog filterbank,
we sampled a bandwidth of 576\,MHz centered on 6.6\,GHz.  Each of 192
polarization-summed frequency channels was sampled every 1\,ms with
one-bit precision.  In parallel, we operated a digital filterbank (DFB)
with a bandwidth of 256\,MHz centered on 6.4\,GHz, which folded the data
into 2048-bin (1\,ms resolution) profiles in all Stokes parameters.

We analyzed the data in standard fashion using the PRESTO, TEMPO, and
PSRCHIVE packages.  The pulsar was detected at $P = 2.0699438(5)$\,s on
MJD~54321.44.  In order to compare the radio and X-ray light curves in
absolute phase, radio pulse times-of-arrival were converted to TDB using
TEMPO, and corrected to infinite frequency using the measured $\mbox{DM}
= 830\pm50$\,cm$^{-3}$\,pc \citep{cam07b}.  At the high radio frequency
used, an uncertainty in DM of 100\,cm$^{-3}$\,pc contributes a negligible
10\,ms uncertainty to the infinite-frequency arrival time.

\begin{figure}[t]
\begin{center}
\hbox{
\hfil
\includegraphics[angle=0,scale=0.475,angle=270.]{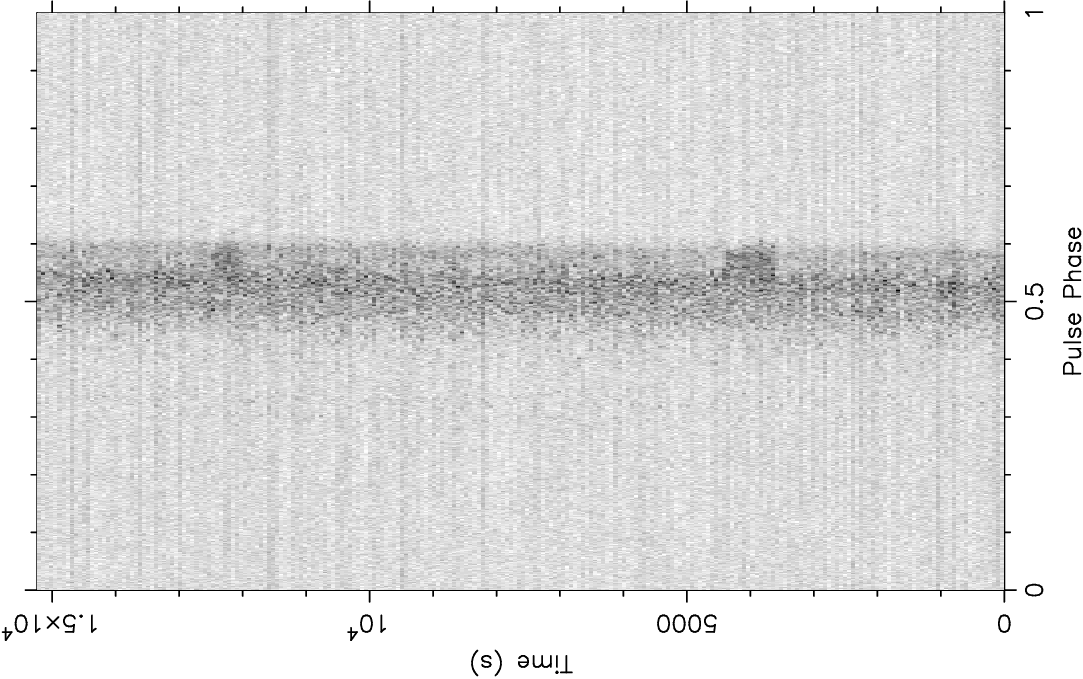}
\includegraphics[angle=0,scale=0.5,angle=270.]{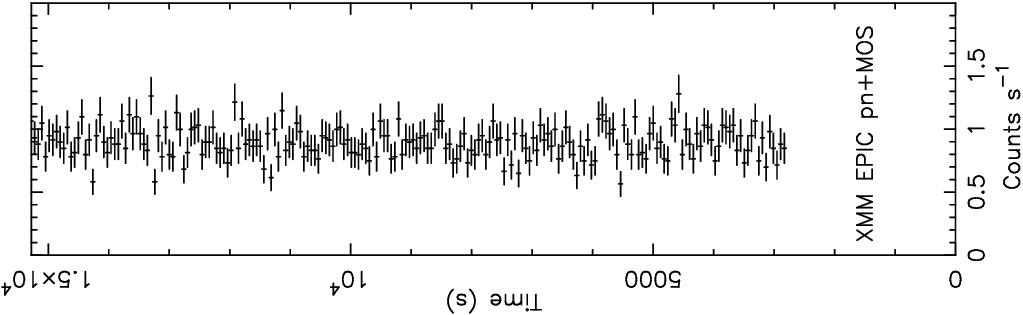}
\hfil
}
\caption{\label{fig:radiopulse}
{\it Left\/}: Radio pulse history obtained with the analog filterbank
during the 4.3 hr Parkes observation
at 6.6 GHz beginning on 2007 August 9 09:57 UT.
Note the transitory
enhancements on the trailing shoulder of the pulse beginning at
3700 s and at 12,000 s.  A fainter event can also be seen at
800 s.
{\it Right\/}: Simultaneous count rate from the \xmm\ EPIC pn and MOS.
}
\end{center}
\end{figure}

The phase alignment is shown in Figure~\ref{fig:radioxraypulse}.
The centroid of the radio pulse lags the X-ray centroid by
0.15 cycles, while the peak of the radio pulse
comes 0.19 cycles after the X-ray centroid.
There were three events in the radio observation in which
the pulse profile suddenly changed for times ranging from
100 to 800 seconds, then returned to its otherwise steady
shape (see Fig.~\ref{fig:radiopulse}).  
This behavior is peculiar to radio emitting AXPs,
and is also seen in \xte\ \citep{cam07a}. No changes
in X-rays are seen in connection with these events,
as the X-ray flux remained constant throughout the
observation.

\section{Discussion}\label{sec:disc} 

\subsection{Distance}

The X-ray-measured $N_{\rm H} \approx 3 \times 10^{22}$\,cm$^{-2}$
exceeds the total Galactic 21~cm \ion{H}{1} column density of
$1.8 \times 10^{22}$~cm$^{-2}$ in the direction of \mag\
\citep{dic90}, suggesting that most of the absorption is due
to molecular material.  The $N_{\rm H}$ equivalent of
the CO column density in this direction is
$N_{\rm H} = 2N_{\rm H_2} \approx 4 \times 10^{22}$~cm$^{-2}$
according to the CO maps of \citet{bro89}.  Half
of the total atomic plus molecular column is then sufficient to
account for the X-ray absorption.

The $\mbox{DM} = 830$\,cm$^{-3}$\,pc
is consistent with the X-ray-measured $N_{\rm H}$
and implies a distance $d \approx 9$\,kpc according to the
free-electron model of \citet{cl02}.
A smaller $d \approx 4$\,kpc was suggested by \citet{gg07} assuming
an association with two nearby star forming regions with \ion{H}{1}
measured distances.  This smaller distance estimate would
place \psr\ in or near the Crux--Scutum spiral arm, while
$d = 9$~kpc would be compatible with a location
near the Norma spiral arm.  At $d \approx 4$\,kpc, the X-ray
luminosities and blackbody areas in Table~\ref{tab:xrayspec}
would be smaller by a factor of 5.

The Galactic longitude of \psr,
$327.24^{\circ}$, falls on the tangent of our line of
sight to the Norma spiral arm as mapped in CO
\citep[see Fig.~9 of][]{bro89}, at a distance of 8.5~kpc.
In this direction the bulk of the CO lies at radial
velocities ranging from $-35$ to $-95$ km~s$^{-1}$.
If \psr\ ever becomes much brighter in the radio,
an \ion{H}{1} absorption study
may be able to pin down its distance more
precisely, since absorption in the Crux-Scutum arm
extends only to negative velocities of $-60$ km~s$^{-1}$,
while velocity as negative as $-90$ km~s$^{-1}$ would
indicate a distance closer to the Norma arm
\citep[see Fig.~4 of][]{mcc01}.

\subsection{Emission Mechanisms and Geometry}

The X-ray spectrum of \mag\
is similar to that of other AXPs, having a thermal
component that even in a low state is fitted by a blackbody of
$kT_{\rm BB} = 0.40$\,keV, which is hotter by at least
a factor of 3 than modeled cooling NSs of its same age
\citep[e.g.,][]{ykhg02}.  This indicates that localized
heating of $\sim 3\%$ of the NS surface area
by magnetic field decay is the primary source
of luminosity in the faintest observed state
of \mag, with a smaller contribution to
the spectrum from magnetospheric scattering.
The remainder of the NS surface is presumably
too cool to be detected, given the large distance and
intervening absorption of soft X-rays.

The 0.15 cycle
offset between the radio and X-ray pulses
recalls the question of whether radio emission from
magnetars arises on open or closed field lines, or
both.  It is possible that magnetic field lines on which
large currents flow and
radio emission is generated are anchored in the area of
concentrated crustal heating from which the thermal
X-ray emission emerges.  In this case, the phase lag
of the radio may by due to the azimuthal twist of
these closed field lines, with the radio emission
coming from higher altitude and different azimuth from the X-rays.
On the other hand, it is not necessary that the X-ray
and radio emission are associated with the same field
lines, especially if radio emission is restricted
to the open field line bundle while X-ray heating
occurs on closed field lines.  The present data
leave this question open.

If we accept the tentative detection of X-ray pulsations in
2006 August, then the pulsed fraction decreased
by a factor of 2 while the pulsed flux
increased by a factor of 8 during the outburst.
Such a trend has been seen in other magnetars, 
notably 1E 1048.1$-$5037 \citep{tie05,gav06}
and CXOU J164710.2$-$455216 \citep{mun07,isr07}.
One possible reason for this effect is the growth of the hot
region, which occupies a significant fraction of the NS
surface in outburst, but not in quiescence.  Another explanation
involves the changing relative contributions of the 
surface blackbody and scattered flux.  A small surface hot spot
may be highly modulated if eclipsed by the rotation of the NS,
which may account for the 100\% pulsed
fraction of CXOU J164710.2$-$455216 in quiescence.  In outburst,
enhanced magnetospheric scattering decreases that modulation.
In the case of \mag, the Comptonized component that we fitted
represents a small fraction ($26-32\%$) of the X-ray flux in
any of its states, so scattering probably does not account for a
change in its pulsed fraction.  However, the increase in
heated area to 16\% of the NS surface may have that effect.

On the other hand, if we discount the evidence of enhanced
pulsed amplitude in 2006, then the small X-ray pulsed
fraction of $\sim 7\%$ in the high state
(by inference in the low state as well),
and the broad radio pulse,
suggest that these emitting regions are both close to 
the axis of rotation, i.e., that \psr\ is nearly
an aligned rotator.  This is in contrast to the other
radio emitting magnetar, \xte.  The pulsed fraction of \xte\
was larger ($\approx 50\%$) at the peak of the outburst
than in quiescence \citep{got04}.
This, plus the polarization properties of the radio pulse,
led \citet{cam07a} to favor large inclination angles for \xte.
The observed characteristics of the
quiescent state of \xte, for which only an upper limit of
$\approx 24\%$ on its pulsed fraction is known \citep{got04},
may also be controlled by its surface thermal emission,
since unlike \mag, the pre-outburst spectrum of \xte\ appears
to be a cool blackbody covering the entire surface,
with no evidence for a surface hot spot.

\subsection{Energetics of the Outburst}

\mag\ resembles the transient AXP \xte\ in that its maximum X-ray observed 
luminosity, $\approx 1.7 \times 10^{35}(d/9\ {\rm kpc})^2$\,ergs\,s$^{-1}$,
is similar to the steady luminosity of most other AXPs, while \mag\ and
\xte\ are usually much fainter.  Assuming that the high state
observed here belongs to a well-defined event, we can extrapolate
the decay to estimate the total integrated energy in the outburst.
Unfortunately, the functional form of the decay is not well established
due to the small dynamic range of the available data, and the
poorly constrained quiescent level.  If we adopt the
linear decay fitted in Figure~\ref{fig:decay} as the
flux to be attributed to the outburst, then
the total energy beginning on 2007 June 22 will be
$\approx 1 \times 10^{42}(d/9\ {\rm kpc})^2$\,ergs.
In comparison, the outburst of \xte\ released
$\approx 6 \times 10^{42}$\,ergs \citep{gh07}.

However, it is likely that the outburst of \mag\ began before our
earliest observation of it.  A serendipitous detection of the
radio pulsar in a survey on 2007 April 23 is reported by S. Johnston
(private communication), from which we infer that the X-ray
outburst was already in progress at least two months prior
to our first \swift\ observation.
The above estimate is then a lower bound
on the outburst energy.  
An upper bound on the total energy can be derived by
hypothesizing that the outburst began shortly after 2006 August 22,
when \mag\ was last observed in a low state.
In this case, the initial
luminosity was 3 times that seen on 2007 June 22,
and the integrated energy is 
$E \approx 1 \times 10^{43}(d/9\ {\rm kpc})^2$\,ergs.

An additional source of uncertainty in the total energy is the
possibly aligned geometry of the pulsar as discussed above.
This could cause us to overestimate the luminosity, which
was calculated assuming isotropic flux, by a factor of a few.
Nevertheless, the estimates are within an order of
magnitude of the energy released in the outburst of \xte.
The latter was interpreted as cooling following a deep crustal
heating event \citep{hal05,guv07}, which is a promising
hypothesis for \mag\ as well.

CXOU J164710.2$-$455216 in Westerlund~1 \citep{isr07,mun07}
is also a transient AXP.  Its quiescent flux was a factor
of $\sim 100$ fainter than its outburst.  The peak luminosity
on the first day of the
outburst was $\approx 1 \times 10^{35}$ ergs~s$^{-1}$.
The flux decay followed a power law of index $-0.28 \pm 0.05$,
and the integrated energy in the outburst was
$\approx 5 \times 10^{41}$ ergs during the first 130 days \citep{isr07},
although it could end up being much higher as the flux
remained well above the quiescent level.
This behavior was similar to the $t^{-0.22}$
decay of the AXP 1E 2259+586 following a bursting episode \citep{woo04}.
The flux integral of such a shallow decay is unbounded, which 
may imply that continuing energy input from the
magnetic field dominates the post-burst decay, rather
than the simple afterglow of an impulsive event.
In contrast to the slow decay of CXOU J164710.2$-$455216
and 1E 2259+586, models of cooling after a deep crustal
heating event follow approximate power-laws in the range $t^{-0.6}-t^{-1.0}$
\citep{lyu02}, while the observed decay of \xte\ was well fitted by an
exponential \citep{gh07}.

Since there was no detection of X-ray bursting activity from
\mag\ prior to its discovery as a radio pulsar, we do not know
if a discrete event was responsible for its current outburst.
Accordingly, it is not yet clear if the decaying X-ray flux is
an afterglow of a sudden injection of energy,
or simply one in a series of fluctuations in its continuous
heating by magnetic field decay.  We do know that earlier
X-ray observations of \mag, by \einstein\ and \asca,
found it at fluxes intermediate between the low and high levels
recorded here \citep{gg07}, suggesting that the recent
outburst may not be an isolated event.  Indeed its X-ray flux
appears to have levelled off, and may even be rising again.

\section{Conclusions}

Prompted by its discovery as a transient radio pulsar,
we obtained new X-ray observations of \psr\
that reveal it in the highest state yet observed, with
$L_X \approx 1.7 \times 10^{35}$
ergs~s$^{-1}$ and declining.
The peak of the outburst was not
observed in X-rays, but
it could have been
several times higher.  As this is the second
AXP to be detected as a transient radio source,
after \xte, we infer that magnetic field rearrangement
accompanying an X-ray outburst may be a necessary
(although not sufficient) condition for radio
turn-on of an AXP.

We also detected X-ray pulsations from
\psr\ for the first time, although the pulsed fraction
$\sim 7\%$ is among the smallest observed from AXPs.
Together with the properties of the simultaneously 
observed radio pulse, this may indicate a nearly
aligned geometry.  In both quiescence and outburst,
the X-ray spectrum is dominated by a small blackbody
hot spot of $kT_{\rm BB} = 0.4-0.5$~keV,
with a lesser Comptonized
component, as is the case for most AXPs.
The changing luminosity is due mostly to the
increase in area of the blackbody, and partly
to its increase in temperature.  This, plus
the energy and time scale of the decay, suggests an origin
of the brightening in a deep crustal heating
event rather than a change in magnetospheric
currents.  The historical X-ray luminosities
of \mag\ span only a factor of 16, including
some intermediate states, which is not as
extreme a range as experienced by \xte\ 
and CXOU J164710.2$-$455216 (a factor of $\sim 100$).
It is not yet clear if we have sampled the total range of
luminosity of \mag, or what the duty cycle and
decay time of its largest outbursts are.

\acknowledgements
This investigation is based on observations obtained with \xmm,
an ESA science mission with instruments and contributions directly
funded by ESA Member States and NASA.
We are grateful to the \swift\ and \xmm\
project scientists and staff for the timely scheduling
of observations.
The Parkes Observatory is part of the Australia Telescope,
which is funded by the Commonwealth of Australia for operation
as a National Facility managed by CSIRO.  This work was
supported by NASA grant NNG05GC43G.

\end{document}